\documentclass[a4paper,11pt]{article}
\pdfoutput=1 

\usepackage{jcappub} 
\usepackage[T1]{fontenc} 
\usepackage{rotating}

\title{\boldmath Constraining CMB physical processes using Planck 2018 data}

\author[a,b]{M. Ruiz-Granda,}
\author[a]{P. Vielva}

\affiliation[a]{Instituto de F\'isica de Cantabria (CSIC-Universidad de Cantabria),\\ 
Avda. de los Castros s/n, E-39005 Santander, Spain}
\affiliation[b]{Dpto. de F\'isica Moderna, Universidad de Cantabria, \\
Avda. los Castros s/n, E-39005 Santander, Spain}

\emailAdd{miguel.ruizgr@alumnos.unican.es}
\emailAdd{vielva@ifca.unican.es}

\abstract{This paper aims to perform a phenomenological parametrisation of the standard cosmological model, $\Lambda$CDM, to weigh the different physical processes that define the pattern of the angular power spectra of the Cosmic Microwave Background (CMB) anisotropies. We use six phenomenological amplitudes to account for the Sachs-Wolfe, early and late Integrated Sachs-Wolfe, polarization contribution, Doppler and lensing effects. To do so, we have adapted the \texttt{CLASS} Boltzmann code and used the Markov Chain Monte Carlo (MCMC) sampler of \texttt{Cobaya} to explore the Planck 2018 likelihood to constrain different combinations of cosmological and phenomenological parameters. Observing deviations of the mean values of the phenomenological amplitudes from the predictions of the $\Lambda$CDM model could be useful to resolve existing cosmological tensions.
For the first time, a comprehensive analysis of the physical processes of the CMB has been performed using the Planck 2018 temperature, polarization and lensing power spectra. In a previous work, the phenomenological amplitudes were constrained using only the TT data, however, by including the polarization and lensing data we find that the constraints on these physical contributions are tighter. In addition, some degeneracies that appear only when considering TT data are completely broken by taking into account all Planck 2018 data. Consequently, models with more than three phenomenological amplitudes can be studied, which is prohibitive when only the temperature power spectrum is used.
The results presented in this paper show that the Planck experiment can constrain all phenomenological amplitudes except the late Integrated Sachs-Wolfe effect. No inconsistencies were found with the $\Lambda$CDM model, and the largest improvements were obtained for the models that include the lensing parameter, $A_L$.
}

\arxivnumber{2206.00731} 

\begin{document}
\maketitle
\flushbottom

\section{Introduction} \label{sec:intro}

The standard cosmological model $\Lambda$CDM studied from the CMB data has allowed the six basic cosmological parameters to be extracted with unprecedented accuracy. The CMB photons, which contain valuable information about the characteristics of our Universe, also contain relevant information about the physical processes they have undergone during the expansion of the Universe. In this work, we study how the introduction of an extra $\Lambda$CDM parametrisation allows us to weigh the different physical processes that define the pattern of the CMB angular power spectra. We define six different phenomenological amplitudes to account for the Sachs-Wolfe, early and late Integrated Sachs-Wolfe, polarization contribution, Doppler and lensing effects. 

This approach is not new, in fact, these physical contributions have been studied in previous works. The early and late Integrated Sachs-Wolfe effects were studied using WMAP and Planck 2015 TT,TE,EE+lowP data in \cite{Cabass_2015}, the lensing effect was measured using Planck 2018 TT,TE,EE+lowE+lensing and also other physical contributions to the temperature spectrum such as the Integrated Sachs-Wolfe, Doppler and Sachs-Wolfe effects were found to be consistent with the $\Lambda$CDM model in \cite{PlanckResults2018}, a comprehensive study of five physical contributions (without the late Integrated Sachs-Wolfe effect) and different combinations of them using only Planck 2018 TT data was presented in \cite{Kable_2020}, and the early Integrated Sachs-Wolfe effect was measured using Planck 2018 temperature and polarization data in \cite{Sunny_2021}.

In this paper, we present a comprehensive analysis of the six different physical contributions mentioned above using the Planck 2018 TT,TE,EE+lowE+lensing data. This study aims to strengthen the constraints for the different phenomenological amplitudes since the polarization and lensing power spectra contain relevant information on the cosmological parameters and physical amplitudes. Another important issue observed in \cite{Kable_2020} was the degeneracy between certain parameters, for example, between the scalar amplitude ($A_s$) and the phenomenological amplitudes for the Sachs-Wolfe effect ($A_{SW}$), the early Integrated Sachs-Wolfe ($A_{eISW}$), and the Doppler ($A_{Dop}$) effects. The full Planck 2018 data allow us to break certain existing degeneracies between the parameters and to study more combinations of the parameters. We address different combinations of the six cosmological parameters and one or more phenomenological amplitudes and compare the obtained results with the Planck TT+lowE 2018 likelihood, which is essentially a replica of the fits made in \cite{Kable_2020} (minor differences explained in detail in section \ref{onePhenomenological}).

For this purpose, we have adapted the equations solved by \texttt{CLASS} \footnote{\url{http://www.class-code.net/}} \cite{Class_citation} Boltzmann code to introduce these new phenomenological parameters. Using \texttt{CLASS} to calculate the theoretical temperature, polarization and lensing angular power spectra, the fits to the Planck 2018 data are performed using the Markov Chain Monte Carlo (MCMC) algorithm adapted from \texttt{CosmoMC} \cite{MCMC_Lewis_2002, MCMC_Lewis_2013, drag_2005} implemented in the \texttt{Cobaya}\footnote{\url{https://ascl.net/1910.019}} software \cite{Cobaya_Torrado_2021}.

This paper is structured as follows. In section \ref{sec:compu}, it is explained how the temperature and polarization angular power spectra are calculated taking into account the phenomenological amplitudes introduced. In addition, different plots are shown to explain the effect of the variation of each phenomenological amplitude on the temperature and polarization power spectra. Then, in section \ref{sec:results} the results of the different MCMC runs for different combinations of cosmological and phenomenological parameters are presented and discussed. Finally, in section \ref{sec:conclusions}, the conclusions of this paper are presented.

\section{Modifications in the calculation of CMB angular power spectra}  \label{sec:compu}
In this section, we present the modifications introduced in the equations for the calculation of the CMB power spectrum in order to weigh the contribution of the different physical effects suffered by the CMB photons.

In this article, only the scalar mode of the perturbations of the metric is studied and, therefore, the different equations are written in terms of the conformal Newtonian gauge. In this situation, the perturbations are characterised by two scalar potentials $\psi$, the Newtonian potential, and $\phi$, the spatial perturbation to the metric, which appear in the line element as:
\begin{equation}
ds^2 = a^2(\eta)[-(1+2\psi(\mathbf{x},\eta))d\tau^2 + (1-2\phi(\mathbf{x},\eta))d\mathbf{x}^2],
\end{equation}
where $\eta$ is the conformal time \cite{Ma_Bertschinger_1995, Dodelson}.

In the framework of the minimal $\Lambda$CDM model, a zero spatial curvature ($K = 0$) is assumed. In this situation, to calculate the scalar power spectra, the following integral must be solved:
\begin{equation}\label{clTheory}
C_\ell^{XY} = 4\pi\int \frac{dk}{k} \Delta_\ell^X(k,\eta_0)\Delta_\ell^Y(k,\eta_0)\mathcal{P}_\mathcal{R}(k),
\end{equation}
where $X,Y \in \{T,E,B\}$, $\Delta_\ell^Y(k,\eta_0)$ are the $Y$ photon transfer function which correspond to the temperature ($T$) or polarization ($E$ or $B$) anisotropies, $\mathcal{P}_\mathcal{R}(k)$ is the primordial curvature power spectrum, $k$ is the wavenumber, and $\eta_0$ is the conformal time today.

Single-field inflation theory predicts a quasi-scale invariant curvature power spectrum, $\mathcal{P}_\mathcal{R}(k)$. To account for deviations from a scale-invariant power spectrum, it is usual to introduce a primordial power-law spectrum:
\begin{equation}
\mathcal{P}_\mathcal{R}(k)=A_sk^{n_s-1},
\end{equation}
where $A_s$ is the amplitude of the scalar spectrum and $n_s$ is called the scalar tilt.

\subsection{Parametrisation of the physical processes affecting the CMB} \label{sec:physical}

The photon transfer functions in the equation \eqref{clTheory} are crucial for weighting the different physical contributions to the CMB angular power spectra. These functions can be obtained by a time integral of a set of source functions multiplied by certain radial functions, as shown in \cite{LineOfSight}. We use the total angular method to calculate these source functions, introduced in \cite{Hu_White_1997}, simplifying the problem of radiation transport under gravity and scattering processes, treated by the Boltzmann equation, for the temperature and polarization anisotropies of the CMB. The total angular momentum method leads to a unified set of simple integrals taking into account the temperature $T$, and the  $E$ and $B$ polarization modes, which allows splitting the source functions into a set of physical contributions as explained in \cite{Fast_CMB_comp_Lesg_2014}. This is the key aspect of this work: modulating with phenomenological amplitudes each physical contribution allows us to weigh them. The resulting photon transfer functions for the scalar modes are:
\begin{equation}\label{transferFunction}
\begin{aligned}
\Delta_\ell^T(k,\eta_0) &=
 A_{SW} \int_{\eta_{ini}}^{\eta_0} d\eta g(\eta)[\Delta^T_0(k,\eta)+\psi(k,\eta)]j_\ell (k(\eta_0-\eta)) +\\
&+\int_{\eta_{ini}}^{\eta_0} d\eta f(\eta)e^{-\tau}(\phi'(k,\eta) + \psi'(k,\eta))j_\ell (k(\eta_0-\eta))+\\
&+A_{Dop}\int_{\eta_{ini}}^{\eta_0} d\eta \frac{g(\eta)\theta_b}{k}j'_\ell (k(\eta_0-\eta))+\\
&+A_{Pol}\int_{\eta_{ini}}^{\eta_0} d\eta \frac{g(\eta)}{2}P^{(0)} [3j''_\ell (k(\eta_0-\eta))+j_\ell (k(\eta_0-\eta))],\\
\Delta_\ell^E(k,\eta_0) &= A_{Pol}\int_{\eta_{ini}}^{\eta_0} d\eta \sqrt{\frac{9(\ell + 2)!}{4(\ell - 2)!}} g(\eta) P^{(0)} \frac{j_\ell (k(\eta_0-\eta))}{(k(\eta_0-\eta))^2},\\
\Delta_\ell^B(k,\eta_0) &= 0,
\end{aligned}
\end{equation}
where $P^{(0)}$ is the scalar polarization source term, $j_\ell (k(\eta_0-\eta))$ are the spherical Bessel functions, $j'_\ell (k(\eta_0-\eta))$ and $j''_\ell (k(\eta_0-\eta))$ are the first and second derivatives of the spherical Bessel functions with respect to their argument, respectively, $\tau$ is the optical depth at a given conformal time, $g(\eta)=-\tau'e^{-\tau}$ is the visibility function, $\Delta^T_0$ is the intrinsic temperature (monopole) anisotropy, and $\theta_b = \nabla\cdot \vec{v}_b$ is the divergence of the baryon bulk velocity, equal to that of the electrons due to the tight Coulomb interactions and associated with the existence of a CMB dipole. In addition, we adopt the following definition for the function that accounts for the early and late ISW regimes:
\begin{equation}
f(\eta)=\left\lbrace 
\begin{aligned}
&A_{eISW} \ \text{for } z \geq 30.\\
&A_{lISW} \ \text{for } z < 30.
\end{aligned}
\right. 
\end{equation}
The reason for choosing the redshift $z=30$ as a turning point is purely phenomenological. As discussed in \cite{Cabass_2015}, the integrand $e^{-\tau}(\phi' + \psi')$ shows a minimum near $z=30$.

The optical depth appearing in equation \eqref{transferFunction}, $\tau = \tau(\eta)$, represents the opacity of the Universe at a given conformal time, $\eta$, when viewed from today, i.e., $\eta=\eta_0$. It is related to the Thomson scattering rate, $\Gamma(\eta)$, by:
\begin{equation}
\tau(\eta) = \int_\eta^{\eta_0}d\eta\Gamma(\eta).
\end{equation}

It is easy to see that it tends to infinity when $\eta\rightarrow 0$, falls below one at recombination, stabilises at a value of the order of 0.1 between the recombination and reionisation epochs, and then decreases smoothly and reaches zero at the present time, by definition. From the CMB angular power spectra, an effective value of $\tau$ can be estimated. The parameter is called the optical depth to reionisation, $\tau_{reio}$, and measures the opacity of the Universe between $z=0$ and $z=z_{reio}$.

The visibility function, $g(\eta)=-\tau'e^{-\tau}$, gives the probability that a CMB photon seen today experienced its last scattering at time $\eta$. It has a narrow and sharp peak around the time of recombination and develops a second, smaller and broader peak around reionisation. This function shows that most of the CMB photons did not interact between the last scattering surface and the current time, while a minority rescattered at the time of reionisation \cite[e.g.,][]{Cosmolo_Lesgourgues_2013}. Connecting this knowledge with equation \eqref{transferFunction} means that the first (Sachs-Wolfe effect), the third (Doppler effect) and the fourth (polarization contribution effect) terms of the temperature photon transfer function originate mainly from the last scattering surface, while the second term (Integrated Sachs-Wolfe effect) originates at all points along the trajectory after recombination (due to the $e^{-\tau}$ factor) since the visibility function does not appear. 

At this point, we have explained all the components that appear in the temperature and polarization transfer functions of equation \eqref{transferFunction}, but we have not yet explained what they represent. The first term, weighted by a phenomenological amplitude $A_{SW}$, comprises the Sachs-Wolfe (SW) effect and includes the intrinsic temperature term $\Delta^T_0$ modified by the gravitational redshift or blueshift of the CMB photons as they leave the last scattering surface due to the effect of the $\psi$ potential. The second term accounts for the Integrated Sachs-Wolfe (ISW) effect, parametrised by $A_{eISW}$ for early times and by $A_{lISW}$ for late times, and contains all the non-conservative effects that occur in a universe with non-static metric fluctuations, e.g., if a potential well becomes shallower in time, photons receive a net blue shift when crossing the well and the CMB appears hotter. The third term accounts for the Doppler shift of CMB photons, modulated by a phenomenological parameter $A_{Dop}$, and produces a gravitational redshift/blueshift from the scattering of moving matter \cite[e.g.,][]{Challinor_2012}. Finally, the fourth term is the contribution of the CMB polarization to the CMB angular power spectra, modulated by the parameter $A_{Pol}$, and is related to the photon polarization and anisotropic Thomson scattering \cite[e.g.,][]{Kable_2020}.

For a complete parametrisation of the physical contributions, it is necessary to consider one last effect: the lensing effect. On their way from the last scattering surface to the present time, CMB photons are deflected by the perturbed gravitational field, which has an imprint on the temperature and polarization angular power spectra \cite[e.g.,][]{RuthDurrer}. In particular, scalar perturbations do not produce a $B$-mode polarization, however, when lensing is included, a non-zero $B$-mode angular power spectrum is expected due to the effect of lensing that partially transforms the $E$-mode into $B$-mode. This lensing contribution is determined by the lensing angular power spectrum, $C_\ell^{\psi}$, which is rescaled by a phenomenological parameter $A_L$ as:
\begin{equation}
\widetilde{C}_\ell^{\psi}= A_LC_\ell^{\psi}.
\end{equation}

The calculation of the lensing angular power spectrum and the corrections introduced in the temperature and polarization power spectra will not be discussed in detail in this paper because those equations have not been modified in \texttt{CLASS}, which follows the full-sky method given in \cite{Challinor_Lewis_2005}. From those equations, we can conclude that the rescaled lensing angular power spectrum is proportional to $A_sA_L$.

If the predictions of $\Lambda$CDM are correct, we would expect to obtain $A_{Pol} = A_{Dop}=A_{eISW}=A_{lISW}=A_{SW}=A_L \equiv 1$, and, conversely, obtaining values that differ from this prediction would indicate inconsistencies with the theory.

\subsection{Physical contributions to the CMB angular power spectra} \label{sec:contributspectrum}

\begin{figure}[ht]
\begin{center}
\includegraphics[width=1.0\textwidth]{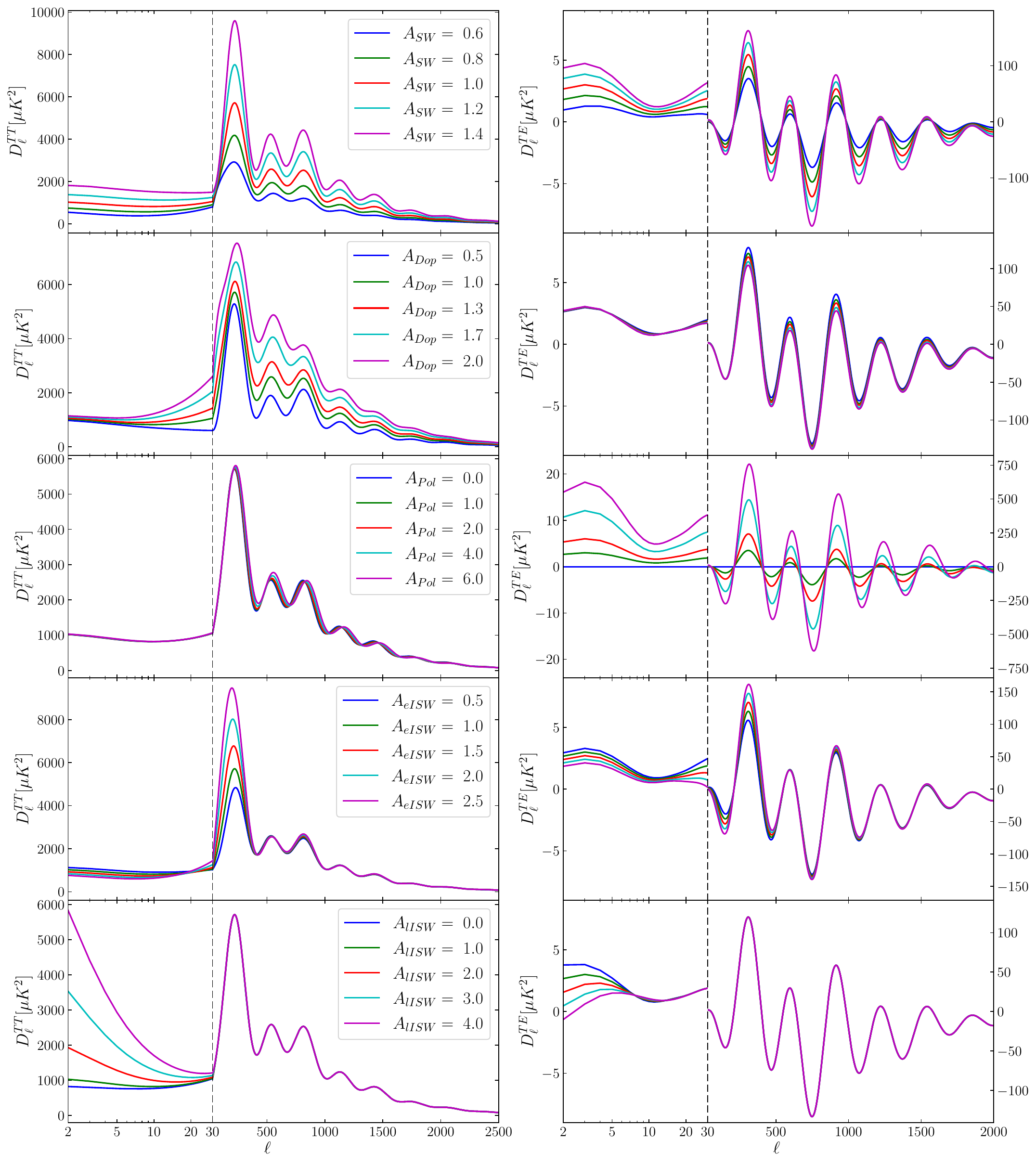}
\caption{\textit{\footnotesize{Plots showing the impact of varying a single physical amplitude at a time on the theoretical TT and TE angular power spectra. The values used for the six cosmological parameters come from the TT,TE,EE+lowE+lensing results of \cite{PlanckResults2018}, and the prediction of the $\Lambda$CDM model corresponds to all phenomenological amplitudes set to one. Note that the scale of the x-axis changes at $\ell=30$, where the horizontal axis changes from logarithmic to linear, which for the TE power spectrum also implies a change in the vertical axis limits.}}}\label{All_latex}
\end{center}
\end{figure}

\begin{figure}[ht]
\begin{center}
\includegraphics[width=1.0\textwidth]{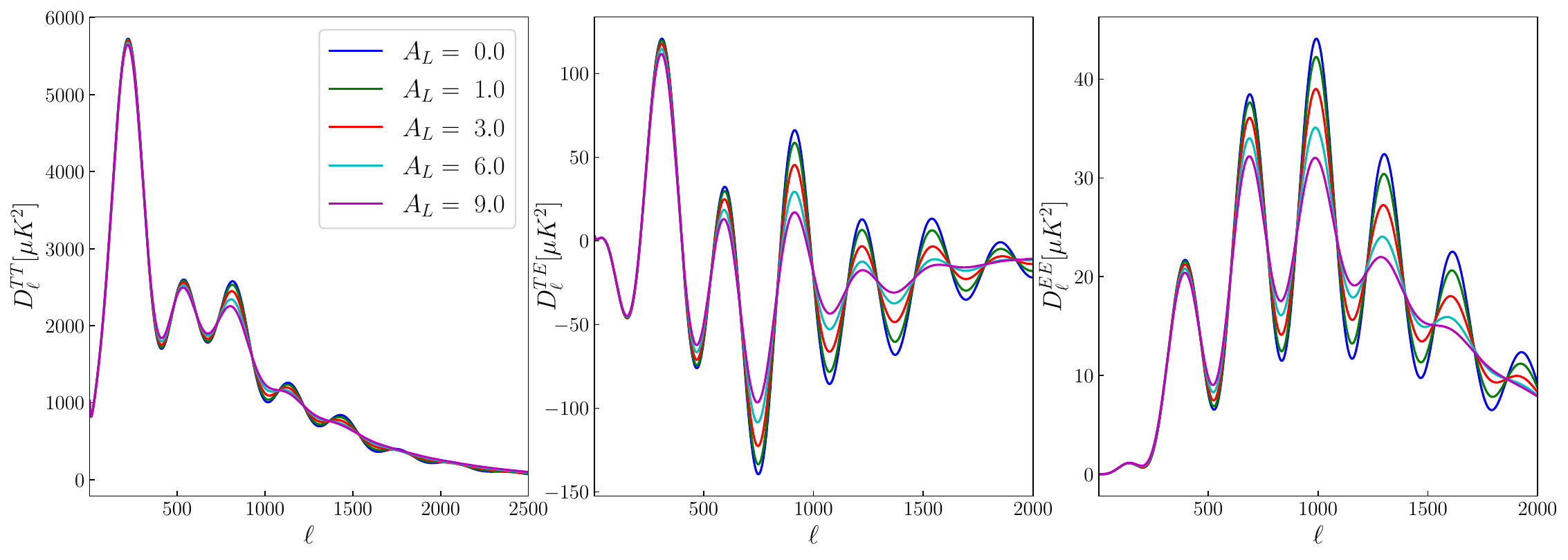}
\caption{\textit{\footnotesize{Plots showing the impact of varying the physical lensing amplitude, $A_L$, on the theoretical TT, EE and TE angular power spectra. The values used for the six cosmological parameters are from the TT,TE,EE+lowE+lensing results of \cite{PlanckResults2018} and the prediction of the $\Lambda$CDM model corresponds to $A_L=1.0$.}}}\label{lensing_latex}
\end{center}
\end{figure}

In this section, we will analyse the effects on the temperature, polarization and lensing angular power spectra of varying each of the phenomenological amplitudes introduced. For this purpose, the \texttt{CLASS} software has been adapted to include the different physical amplitudes introduced in section \ref{sec:physical}. In Figure \ref{All_latex} the effects of varying the phenomenological amplitude corresponding to the Sachs-Wolfe (SW), the early (eISW) and late (lISW) Integrated Sachs-Wolfe, the polarization contribution (Pol) and the Doppler effects (Dop) on the TT and TE power spectra can be observed. Figure \ref{lensing_latex} shows the effects of the variation of the lensing phenomenological amplitude, $A_L$, on the TT, TE and EE power spectra. In both figures, the $y$-axis corresponds to $D_\ell=\ell(\ell+1)C_\ell/2\pi$.

From equation \eqref{transferFunction} it can be seen that, except for the lensing and polarization contribution effects, the rest of the phenomenological amplitudes do not affect the $E$-mode polarization power spectra. Moreover, although the effect of the polarization contribution has an imprint on the EE power spectrum, it is not depicted because it only scales $C_\ell^{EE}$ by a factor proportional to $A_{Pol}$. Finally, we note that the lensing angular power spectrum is only affected by the phenomenological amplitude $A_L$, which scales the spectrum, so it is not interesting to represent it either.

One of the main objectives of this work is to show that the use of CMB polarization data is advantageous to obtain a better constraint of the phenomenological amplitudes. The cross-correlation between temperature and $E$-mode polarization (TE) is strongly affected by variations of these new parameters. Another motivation for including polarization and lensing data comes from the discussion in \cite{Kable_2020}, where it is shown that the scalar amplitude $A_s$ is highly anti-correlated with the combination of the most dominant contributions to the temperature angular power spectrum: Sachs-Wolfe, Doppler and early Integrated Sachs-Wolfe. As will be shown in section \ref{sec:results}, the inclusion of the polarization data permits to break of the existing degeneracies between these parameters. 

Focusing now on Figure \ref{All_latex}, the Sachs-Wolfe effect corresponds to the two plots in the first row. As can be seen, changing $A_{SW}$ affects the entire TT and TE angular power spectrum. In particular, increasing the parameter $A_{SW}$ increases both peaks and valleys in the TT power spectrum, while for the TE angular power spectrum peaks increase and valleys decrease. The second row of plots corresponds to the Doppler effect. This effect is important at intermediate and small scales ($\ell \geq 10$). It increases both peaks and valleys in the temperature power spectrum, but not to the same extent. In the TE angular power spectrum, the peaks increase and the valleys decrease.

The effect of varying the $A_{Pol}$ parameter is also shown in Figure \ref{All_latex}. The effect of the polarization contribution produces a phase shift in both the TT and TE angular power spectra, mainly affecting the second and following peaks. In addition, this effect raises the temperature and polarization power spectra when $A_{Pol}$ is increased, especially for the EE and TE angular power spectra, since $A_{Pol}$ acts as a multiplicative constant of the spectra in these two situations. The fourth row corresponds to the early Integrated Sachs-Wolfe effect, which has its greatest impact on the first peak in both TT and TE. In both cases the height of the first peak increases as $A_{eISW}$ increases. Finally, in the last row, the effect of varying the $A_{lISW}$ parameter is shown. The late Integrated Sachs-Wolfe effect has its largest impact at large scales ($A_{lISW}$) in both the TT and TE power spectra. It increases the height of the TT angular power spectrum and decreases the height of the TE angular power spectrum at low  $\ell$  when $A_{lISW}$ is increased.

\newpage

Finally, in Figure \ref{lensing_latex} the effects of varying the $A_{L}$ parameter are shown. The lensing effect smooths the TT, EE and TE angular power spectra, especially in small scales, and multiplies by a factor $A_L$ the lensing power spectrum.

\section{Results} \label{sec:results}
In this section, we explain the methodologies used and present the constraints for different extra-parametrisations of the $\Lambda$CDM model, including one to four phenomenological amplitudes. A thorough analysis of both the cosmological parameters and the new phenomenological parameters would allow us to determine whether the $\Lambda$CDM model is compatible with the CMB observations or, on the contrary, there might be some new physics hidden. To pursue this goal, it will not only be important to focus on the estimated mean values and parameter uncertainties but also on the goodness of fit (given in terms of $\chi^2$) of the different cases. In addition, another key aspect of our work is to study the degeneracies that arise between the cosmological and phenomenological parameters when using the CMB data. In this way, it will be possible to identify the constraints for fitting the parameter combinations of the different models using only the CMB data.

The methodological approach followed in this work is explained below. The different settings presented in this work are performed using a Markov Chain Monte Carlo (MCMC) sampler from \texttt{Cobaya} \cite{Cobaya_Torrado_2021} based on the software \texttt{CosmoMC} \cite{MCMC_Lewis_2002,MCMC_Lewis_2013,drag_2005} used for Planck likelihood analysis. The sampler requires a theory code, which calculates the CMB angular power spectra for a given set of parameters, priors, which are defined for each parameter as flat priors, and a likelihood, which essentially contains the CMB data to be fitted and is provided in the Planck Legacy Archive \cite{Likelihood_Planck_2018}.

For the theory code, as explained in three previous sections we have adapted the photon transfer functions calculated by \texttt{CLASS} \cite{Class_citation} Boltzmann code to introduce the new phenomenological parameters.

For the likelihood, the latest data from the Planck mission are used. We used different Planck 2018 likelihoods: plikHM\_TT (TT spectrum only, $30\leq \ell\leq 2508$), plikHM\_TTTEEE (TT+TE+EE joint constraint, $30\leq \ell\leq 2508$), lowl (low-$\ell$ Planck temperature, $2\leq \ell\leq 29$, from \texttt{Commander}), lowE (low-$\ell$ HFI EE polarization only from \texttt{SimAll}, $2\leq \ell\leq 29$) \cite{Likelihood_Planck_2018} and lensing (Planck lensing power spectrum reconstruction) \cite{Lensing_Planck_2018}. The low $\ell$ TE power spectrum is not considered in this paper because it has been excluded by the Planck Collaboration in the 2018 release. As explained in \cite{PlanckResults2018}, the cause of the discard is the existence of an excess of variance compared to simulations at low multipoles due to not yet well-understood systematics. In particular, two different sets of likelihoods are used throughout this work: the temperature-only likelihood, and the temperature, polarization and lensing likelihood. The first likelihood is the combination of TT+lowl+lowE, in which the low multipoles of the EE power spectrum are included with the main goal of having a good constraint on $\tau$ and avoiding the introduction of a strong prior on this parameter. Hereafter, this combination will be called TT+lowE. The second likelihood is the combination of plikHM\_TTTEEE+lowl+lowE+lensing, where the full power of Planck mission is used, and which we will name as TT,TE,EE+lowE+lensing from now on. 

The MCMC algorithm implemented in \texttt{Cobaya} uses a generalised version of the $R-1$ Gelman-Rubin statistic \cite{GelmanRubin} to check convergence. In this paper, the stop value of $R-1$ is set to 0.05 for all runs performed. The output of the MCMC is a collection of samples obtained through a random walk in parameter space. For model comparison, the minimum value of $\chi^2$ was estimated for each run using a minimisation sampler \cite{minimize1, minimize2, minimize3} run after the MCMC algorithm was executed. To analyse these samples, a Python package called \texttt{GetDist}\footnote{\url{https://getdist.readthedocs.io/}} \cite{LewisGetDist} has been used. This program allows quantities of interest to be calculated from the samples, such as parameter means, confidence intervals and marginal densities. With \texttt{GetDist} it is possible to obtain 2D contour plots with 68\% and 95\% of the samples, which helps us to study degeneracies between parameters.

\subsection{One phenomenological amplitude} \label{onePhenomenological}

In this subsection, we aim to extend the results presented in \cite{Kable_2020} by testing an additional phenomenological amplitude of the late Integrated Sachs-Wolfe effect, $A_{lISW}$, and analyse whether tighter constraints can be achieved by including the Planck 2018 polarization and lensing data.

Unlike the approach followed in \cite{Kable_2020}, here we do not use a lite likelihood version, where the foreground parameters have already been marginalised. Instead, we allow these parameters to vary and check whether they have any impact on the cosmological and phenomenological parameters when compared to the results in \cite{Kable_2020}. Moreover, as explained above, in $\tau$ no strong prior is introduced, and this parameter will be mainly constrained by the inclusion of low multipole EE data. These two minor aspects make our analysis different and act as a consistency check with the results presented in \cite{Kable_2020}.

\begin{figure}[ht]
\begin{center}
\includegraphics[width=1.0\textwidth]{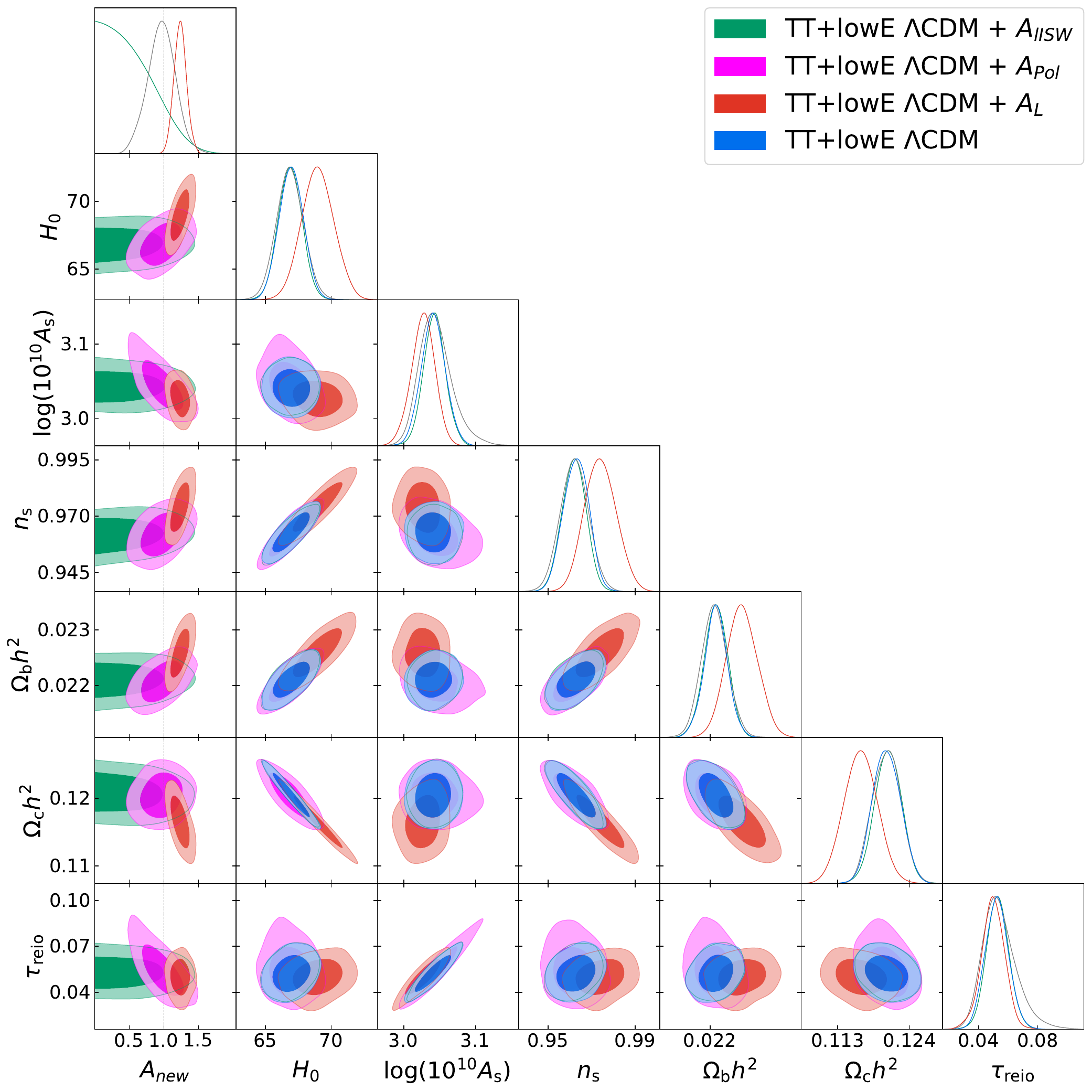}
\caption{\textit{\footnotesize{Plots of the marginalized posterior comparison of $\Lambda$CDM + $A_{lISW}$, $\Lambda$CDM + $A_{Pol}$, $\Lambda$CDM + $A_{L}$ and $\Lambda$CDM fits to Planck 2018 TT+lowE data. The contours display the 68\% and 95\% limits and the black dotted line represent the value for each phenomenological amplitude, $A_{new}=1$, predicted by $\Lambda$CDM model.}}}\label{TTlISWPolLens}
\end{center}
\end{figure}

\begin{figure}[ht]
\begin{center}
\includegraphics[width=1.0\textwidth]{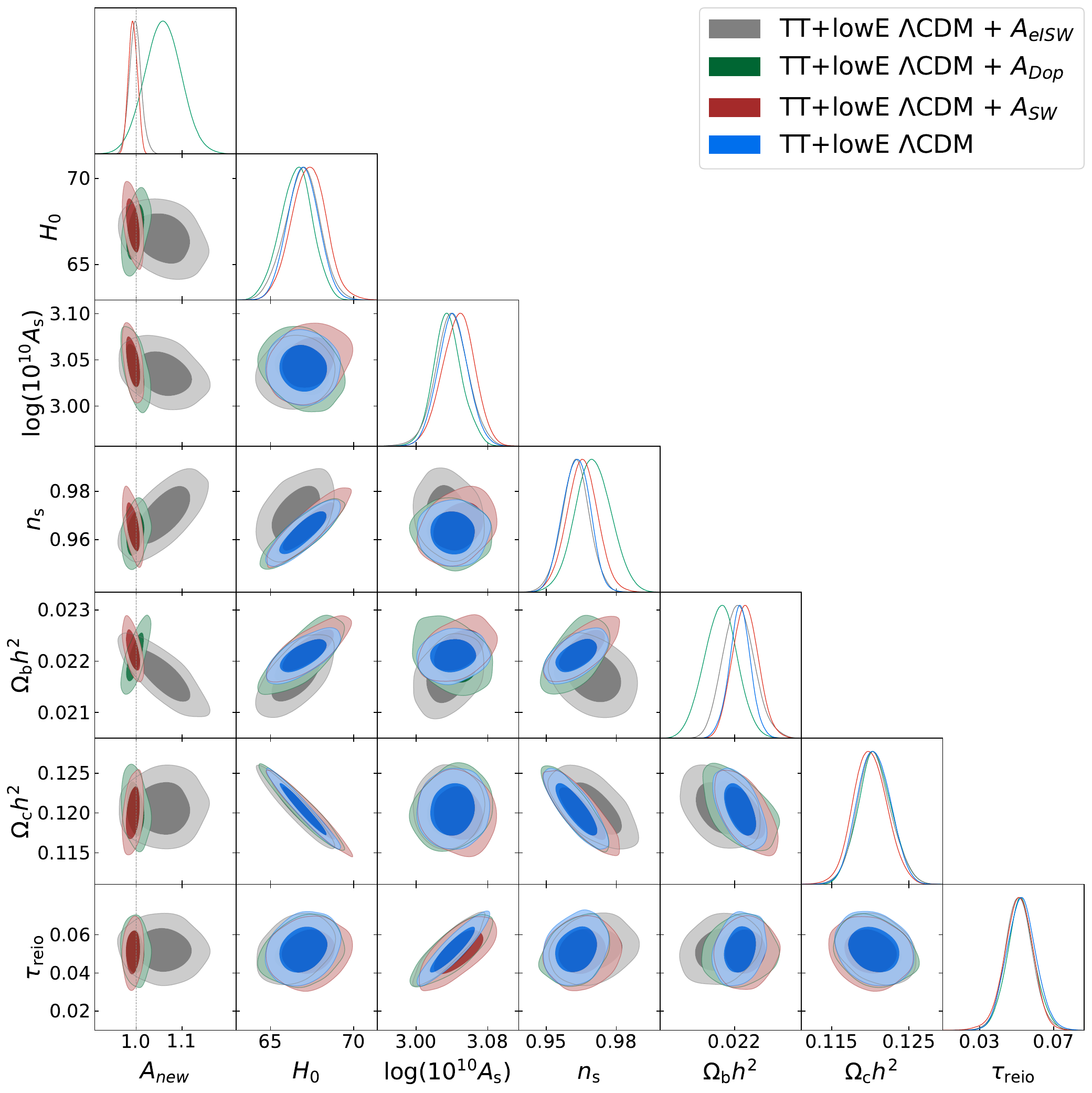}
\caption{\textit{\footnotesize{Plots of the marginalized posterior comparison of $\Lambda$CDM + $A_{eISW}$, $\Lambda$CDM + $A_{Dop}$, $\Lambda$CDM + $A_{SW}$ and $\Lambda$CDM fits to Planck 2018 TT+lowE data. The contours display the 68\% and 95\% limits and the black dotted line represent the value for each phenomenological amplitude, $A_{new}=1$, predicted by $\Lambda$CDM model.}}}\label{TTeISWDopSW}
\end{center}
\end{figure}

\begin{figure}[ht]
\begin{center}
\includegraphics[width=1.0\textwidth]{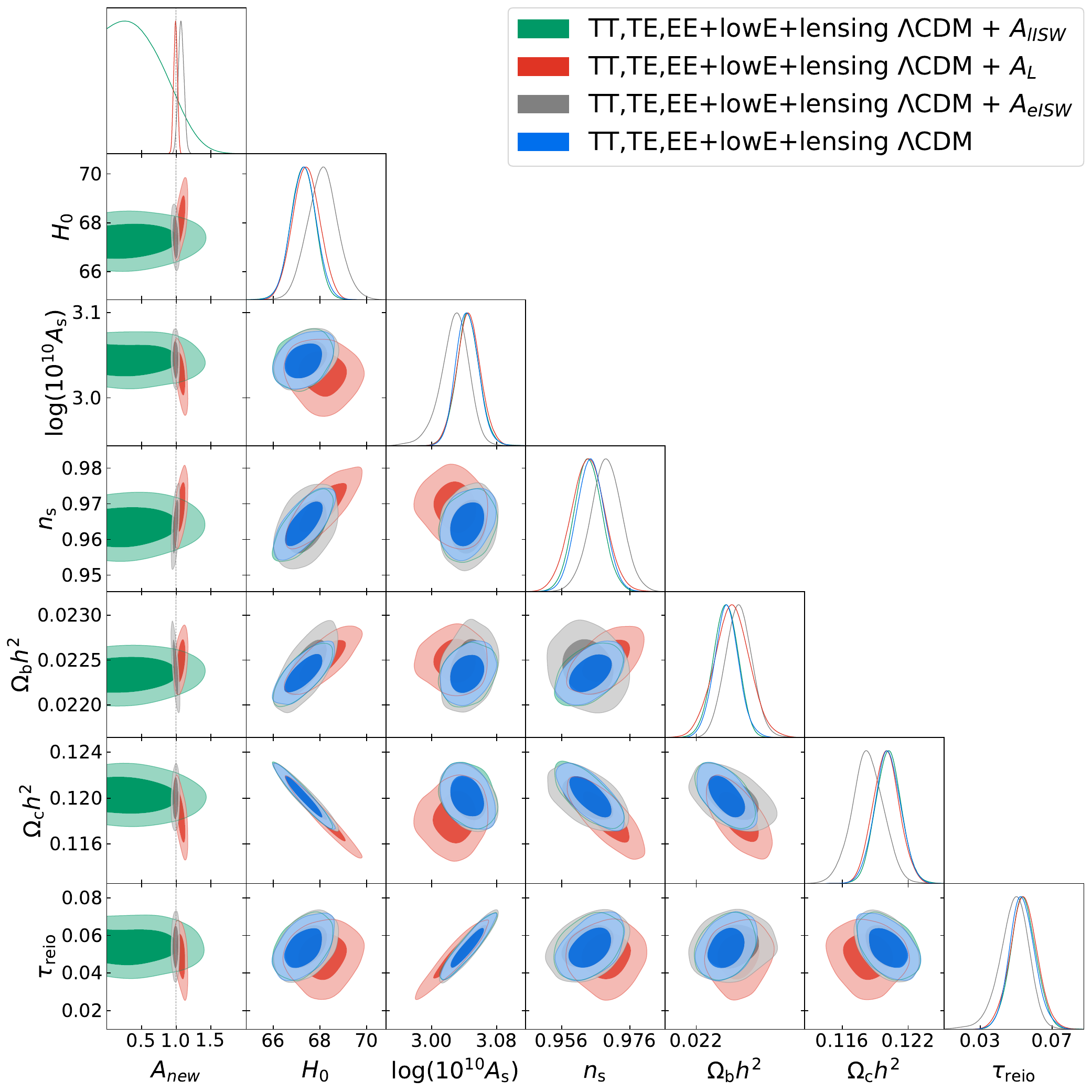}
\caption{\textit{\footnotesize{Plots of the marginalized posterior comparison of $\Lambda$CDM + $A_{lISW}$, $\Lambda$CDM + $A_{L}$, $\Lambda$CDM + $A_{eISW}$ and $\Lambda$CDM fits to Planck 2018 TT,TE,EE+lowE+lensing data. The contours display the 68\% and 95\% limits and the black dotted line represent the value for each phenomenological amplitude, $A_{new}=1$, predicted by $\Lambda$CDM model.}}}\label{TTTEEEeISWlISWLens}
\end{center}
\end{figure}

\begin{figure}[ht]
\begin{center}
\includegraphics[width=1.0\textwidth]{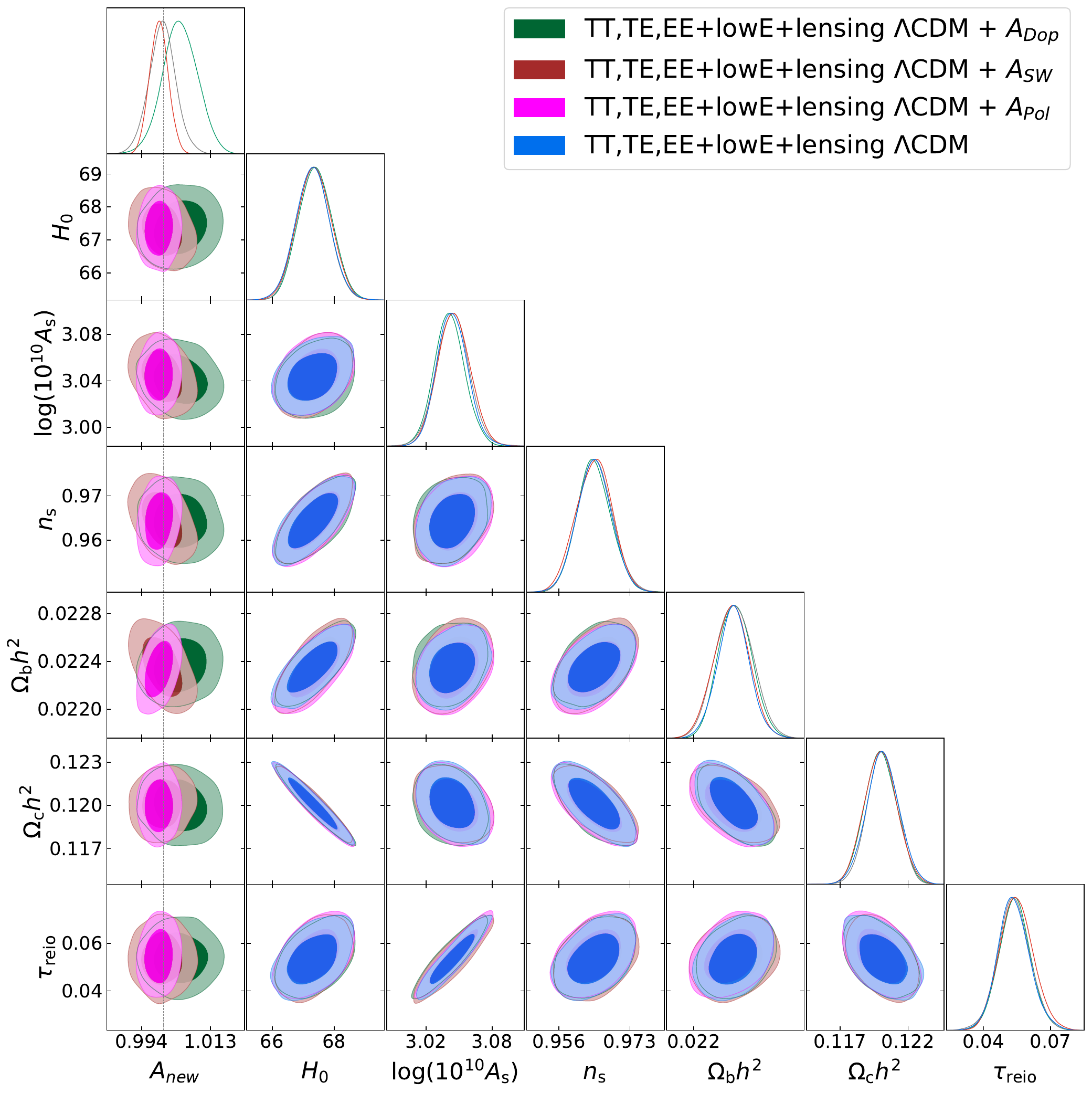}
\caption{\textit{\footnotesize{Plots of the marginalized posterior comparison of $\Lambda$CDM + $A_{Dop}$, $\Lambda$CDM + $A_{SW}$, $\Lambda$CDM + $A_{Pol}$ and $\Lambda$CDM fits to Planck 2018 TT,TE,EE+lowE+lensing data. The contours display the 68\% and 95\% limits and the black dotted line represent the value for each phenomenological amplitude, $A_{new}=1$, predicted by $\Lambda$CDM model.}}}\label{TTTEEEDopSWPol}
\end{center}
\end{figure}

In Tables \ref{tab:TTOnePhenom} and \ref{tab:TTPolOnePhenom} the mean and 68\% confidence intervals are shown for the different cosmological and phenomenological parameters obtained from the Planck 2018 TT+lowE and TT,TE,EE+lowE+lensing data, respectively. In Figures \ref{TTlISWPolLens} and \ref{TTeISWDopSW} the results obtained using Planck 2018 TT+lowE data are shown, while in Figures \ref{TTTEEEeISWlISWLens} and \ref{TTTEEEDopSWPol} the results using Planck 2018 TT,TE,EE+lowE+lensing data are presented.

The phenomenological amplitudes best constrained using the TT+lowE data are the Sachs-Wolfe effect, $A_{SW}$, and the Doppler effect, $A_{Dop}$. This is easily understood from section \ref{sec:contributspectrum} because these effects have the largest impact on the CMB temperature angular power spectrum. The worst-determined amplitude is the late Integrated Sachs-Wolfe effect, $A_{lISW}$, which remains unconstrained because it dominates at low $\ell$ where the cosmic variance imposes a severe limitation. It is well known that the lISW effect cannot be detected using only the CMB data (as shown in Figures \ref{TTlISWPolLens} and \ref{TTTEEEeISWlISWLens}) and cross-correlation with other data sets is needed. In fact, a 4$\sigma$ detection was obtained from the joint cross-correlation of the Planck CMB data with different large-scale structure (LSS) tracers, such as radio sources from the NVSS catalogue, galaxies from the optical SDSS and the infrared WISE surveys \cite{ISWeffectPlanck}. Notice, however, that values of the $A_{lISW}$ parameter smaller than one are favoured, which could be related to the known lack of power of the TT power spectrum on the largest scales.

The second worst constrained phenomenological parameter is $A_{Pol}$ because it has little effect on the temperature angular power spectrum. Finally, the early Integrated Sachs-Wolfe effect and the lensing effect are determined with almost the same uncertainty. However, while the rest of phenomenological amplitudes are compatible with a value of 1 at $1\sigma$ confidence interval, for the $A_{eISW}$ parameter a $1.5\sigma$ deviation is observed, and for the $A_L$ parameter a shift of $2.6\sigma$ to its expected value is reported, which was reported in \cite{PlanckResults2018} by Planck Collaboration.

We are also interested in the correlation of the phenomenological amplitudes with the scalar amplitude, $A_s$, which could be degenerated as it rescales the entire temperature and polarization angular power spectrum. When considering only the temperature data, the largest anti-correlation is obtained for the effect of the polarization contribution due to the introduction of lowE data. The underlying reason for this result is probably related to equation \eqref{transferFunction}, where the introduction of $A_{Pol}$ in the E-mode polarization transfer function implies that it is rescaled by $A_{Pol}^2$ and $A_s$, introducing a strong degeneracy. Considering that only 28 lowE data points are available, the effect produced by the polarization contribution effect in the TT spectrum is so small that it is unable to break the existing degeneracy in the EE angular power spectrum.

The second and third most anti-correlated phenomenological amplitudes with $A_s$ are the Doppler effect and the Sachs-Wolfe effect because they essentially produce an overall effect similar to scaling the entire temperature angular power spectrum.  Finally, to a lesser extent, the eISW effect is also anti-correlated, because it scales the first peak of the TT spectrum, and the lensing effect as well, because the lensing angular power spectrum is proportional to $A_L$ and $A_s$ \cite{Challinor_Lewis_2005}. No correlation with $A_{lISW}$ was observed. 

When polarization and lensing data are included, all phenomenological parameters are better constrained. One of the causes could be the reduction of about 50\% in the uncertainty of $A_s$ because the angular power spectrum of EE is rescaled by $A_{Pol}^2$ and $A_s$. Therefore, there is a degeneracy between $A_{Pol}$ and $A_{s}$ when only the EE angular power spectrum is taken into account, but it breaks down completely when the TT spectrum is also considered since the effect of the polarization contribution shifts the position of the peaks and does not rescale the spectrum as $A_s$ does. Therefore, in this situation, the effect of the polarization contribution modulated by $A_{Pol}$ goes from being one of the worst constrained parameters to being the best of the six phenomenological amplitudes considered in this work. As for the late Integrated Sachs-Wolfe effect, its main imprint on the polarization-related spectra affects the low $\ell$ region of the TE angular power spectrum. Unfortunately, as already discussed, this spectrum was not included in the 2018 version of likelihood. Moreover, as discussed above, the detection of this effect is highly dependent on the CMB and large-scale structure correlations. Note that in \cite{Cabass_2015}, they were not able to constrain the lISW effect even using the full Planck 2015 likelihood (without lensing), where the lowP likelihood included the low $\ell$ TE angular power spectrum. Therefore, for all these reasons, there is no real constraint on this parameter.

In Table \ref{tab:TTPolOnePhenom} we can observe that all phenomenological amplitudes are compatible with one with a confidence interval of $1\sigma$, except for $A_L$. The $1.5\sigma$ deviation observed for $A_{eISW}$ when using the TT+lowE data is reduced to less than $1\sigma$ when using the full Planck 2018 likelihood. For the lensing amplitude, we observe a considerable reduction from $2.6\sigma$ to $1.8\sigma$ from the expected value of one.

The anti-correlation effect observed for parameters $A_{SW}$, $A_{Dop}$ and $A_{eISW}$ is still present when polarization and lensing data are considered, but it is reduced notably for these parameters. However, $A_L$ is highly anti-correlated with the scalar amplitude because the lensing angular power spectrum is proportional to $A_s A_L$. Both $A_{Pol}$ and $A_{lISW}$ are practically not correlated with $A_s$.

\begin{table}[htbp]
  \centering
  \resizebox{\columnwidth}{!}{%
    \begin{tabular}{|c|c|c|c|c|c|c|c|}
    \hline
    Param. & $\Lambda$CDM & $+A_{SW}$ & $+A_{eISW}$ & $+A_{lISW}$ & $+A_{L}$ & $+A_{Dop}$ & $+A_{Pol}$  \\
    \hline\hline
    $n_s$ & $0.9629\pm 0.0057$ & $0.9654\pm 0.0066$ & $0.9701\pm 0.0077$ & \multicolumn{1}{l|}{$0.9620\pm 0.0055$} & $0.9741\pm 0.0072$ & $0.9626\pm 0.0059$ & $0.9621\pm 0.0064$  \\
    \hline
    $H_0$ & $66.96\pm 0.92$ & $67.3\pm 1.0$ & $66.55\pm 0.96$ & $66.87\pm 0.90$ & $69.0\pm 1.2$ & $66.9\pm 1.0$ & $66.9\pm 1.0$  \\
    \hline
    $100\Omega_bh^2$ & $2.211\pm 0.022$ & $2.222\pm 0.026$ & $2.172\pm 0.033$ & $2.211\pm 0.022$ & $2.259\pm 0.029$ & $2.208\pm 0.031$ & $2.208\pm 0.024$  \\
    \hline
    $\Omega_ch^2$ & $0.1205\pm 0.0021$ & $0.1200\pm 0.0022$ & $0.1206\pm 0.0021$ & $0.1207\pm 0.0020$ & $0.1165\pm 0.0025$ & $0.1205^{+0.0020}_{-0.0023}$ & $0.1205\pm 0.0021$  \\
    \hline
    $\tau_{reio}$ & $0.0525\pm 0.0079$ & $0.0509^{+0.0079}_{-0.0071}$ & $0.0522\pm 0.0075$ & $0.0529\pm 0.0076$ & $0.0498\pm 0.0080$ & $0.0512\pm 0.0075$ & $0.0544^{+0.0087}_{-0.013}$  \\
    \hline
    $\ln (10^{10}A_s)$ & $3.041\pm 0.017$ & $3.047\pm 0.018$ & $3.035^{+0.015}_{-0.017}$ & $3.043\pm 0.015$ & $3.027^{+0.017}_{-0.015}$ & $3.040\pm 0.018$ & $3.045^{+0.018}_{-0.025}$  \\
    \hline
    $A_{new}$ & $-$   & $0.9929\pm 0.0098$ & $1.058\pm 0.040$ & $< 0.680$ & $1.235\pm 0.091$ & $0.998\pm 0.013$ & $0.97\pm 0.20$  \\
    \hline\hline
    $\chi^2$ & 1172.18 & 1172.42 & 1170.96 & 1170.33 & 1166.82 & 1172.16 & 1173.79 \\
    \hline
    $\chi^2_{\Lambda CDM} - \chi^2$ & 0.00     & -0.24 & 1.22  & 1.85  & 5.36  & 0.02  & -1.61 \\
    \hline
    \end{tabular}%
    }
    \caption{Mean values and 68\% confidence intervals for the minimal $\Lambda$CDM parameters plus one phenomenological amplitude for the MCMC chains fit to Planck 2018 TT+lowE data.}
  \label{tab:TTOnePhenom}%
\end{table}%

\begin{table}[htbp]
  \centering
  \resizebox{\columnwidth}{!}{
    \begin{tabular}{|c|c|c|c|c|c|c|c|}
    \hline
    Param. & $\Lambda$CDM & $+A_{SW}$ & $+A_{eISW}$ & $+A_{lISW}$ & $+A_{L}$ & $+A_{Dop}$ & $+A_{Pol}$  \\
    \hline \hline
    $n_s$ & $0.9645\pm 0.0041$ & $0.9646\pm 0.0041$ & $0.9637\pm 0.0049$ & $0.9638\pm 0.0041$ & $0.9691\pm 0.0046$ & $0.9644\pm 0.0040$ & $0.9644\pm 0.0043$  \\
    \hline
    $H_0$ & $67.30\pm 0.54$ & $67.32\pm 0.52$ & $67.41\pm 0.57$ & $67.28\pm 0.52$ & $68.12\pm 0.66$ & $67.39\pm 0.53$ & $67.35\pm 0.54$  \\
    \hline
    $100\Omega_bh^2$ & $2.235\pm 0.014$ & $2.236\pm 0.016$ & $2.242\pm 0.021$ & $2.234\pm 0.014$ & $2.250\pm 0.016$ & $2.237\pm 0.015$ & $2.234\pm 0.016$  \\
    \hline
    $\Omega_ch^2$ & $0.1201\pm 0.0012$ & $0.1201\pm 0.0011$ & $0.1200\pm 0.0012$ & $0.1202\pm 0.0012$ & $0.1183\pm 0.0014$ & $0.1200\pm 0.0012$ & $0.1200\pm 0.0012$  \\
    \hline
    $\tau_{reio}$ & $0.0536^{+0.0065}_{-0.0074}$ & $0.0537\pm 0.0073$ & $0.0542\pm 0.0077$ & $0.0541\pm 0.0073$ & $0.0488^{+0.0089}_{-0.0070}$ & $0.0540\pm 0.0070$ & $0.0547\pm 0.0075$  \\
    \hline
    $\ln (10^{10}A_s)$ & $3.044\pm 0.014$ & $3.044\pm 0.015$ & $3.045\pm 0.015$ & $3.045\pm 0.014$ & $3.028^{+0.019}_{-0.015}$ & $3.041\pm 0.014$ & $3.045\pm 0.015$  \\
    \hline
    $A_{new}$ & $-$   & $0.9997\pm 0.0037$ & $0.991\pm 0.026$ & $< 0.687$ & $1.072\pm 0.040$ & $1.0046\pm 0.0048$ & $0.9988\pm 0.0026$  \\
    \hline \hline
    $\chi^2$ & 2766.81 & 2764.81 & 2766.28 & 2764.83 & 2764.93 & 2766.94 & 2765.05 \\
    \hline
    $\chi^2_{\Lambda CDM} - \chi^2$ & 0.00     & 2.00     & 0.53  & 1.98  & 1.88  & -0.13 & 1.76 \\
    \hline
    \end{tabular}%
    }
    \caption{Mean values and 68\% confidence intervals for the minimal $\Lambda$CDM parameters plus one phenomenological amplitude for the MCMC chains fit to Planck 2018 TT,TE,EE+lowE+lensing data.}
  \label{tab:TTPolOnePhenom}%
\end{table}%

At this point, it is interesting to analyse the $\chi^2$ of the different settings. In particular, we are interested in studying whether an improvement is observed or not, compared to the $\Lambda$CDM setting. Thus, the Tables \ref{tab:TTOnePhenom} and \ref{tab:TTPolOnePhenom} both have a row,  $\chi^2_{\Lambda CDM} - \chi^2$, which basically indicates an improvement of the $\Lambda$CDM fit if the obtained value is positive enough, according to the number of extra parameters. For the TT+lowE fits, the largest improvement is observed for the model with the lensing parameter, followed by the model with the early ISW parameter. The underlying reason is that the $A_L$ parameter models the oscillatory residual at high-$\ell$ TT data \cite{PlanckResults2018}. When polarization and lensing data are included, the preferred model is $A_{SW}$, which has only a slightly smaller $\chi^2$ than $A_L$. The lISW effect is not constrained, so its $\chi^2$ value has not been taken into consideration in this discussion.

Checking the results obtained in Table \ref{tab:TTOnePhenom} with \cite{Kable_2020}, an excellent agreement between them has been observed in terms of 68\% confidence intervals. Other models were compared with existing bibliography, such as $A_L$ from \cite{PlanckResults2018} and the 2018 cosmological parameters and MC chains at the Planck Legacy Archive\footnote{\url{https://wiki.cosmos.esa.int/planck-legacy-archive/index.php/Cosmological_Parameters}}, and $A_{eISW}$ and $A_{lISW}$ from \cite{Cabass_2015}. For those cases, a good agreement was found as well. 

\clearpage
\subsection{Two and three phenomenological amplitudes} \label{twoThreeParamsSection}

\begin{figure}[ht]
\begin{center}
\includegraphics[width=1.0\textwidth]{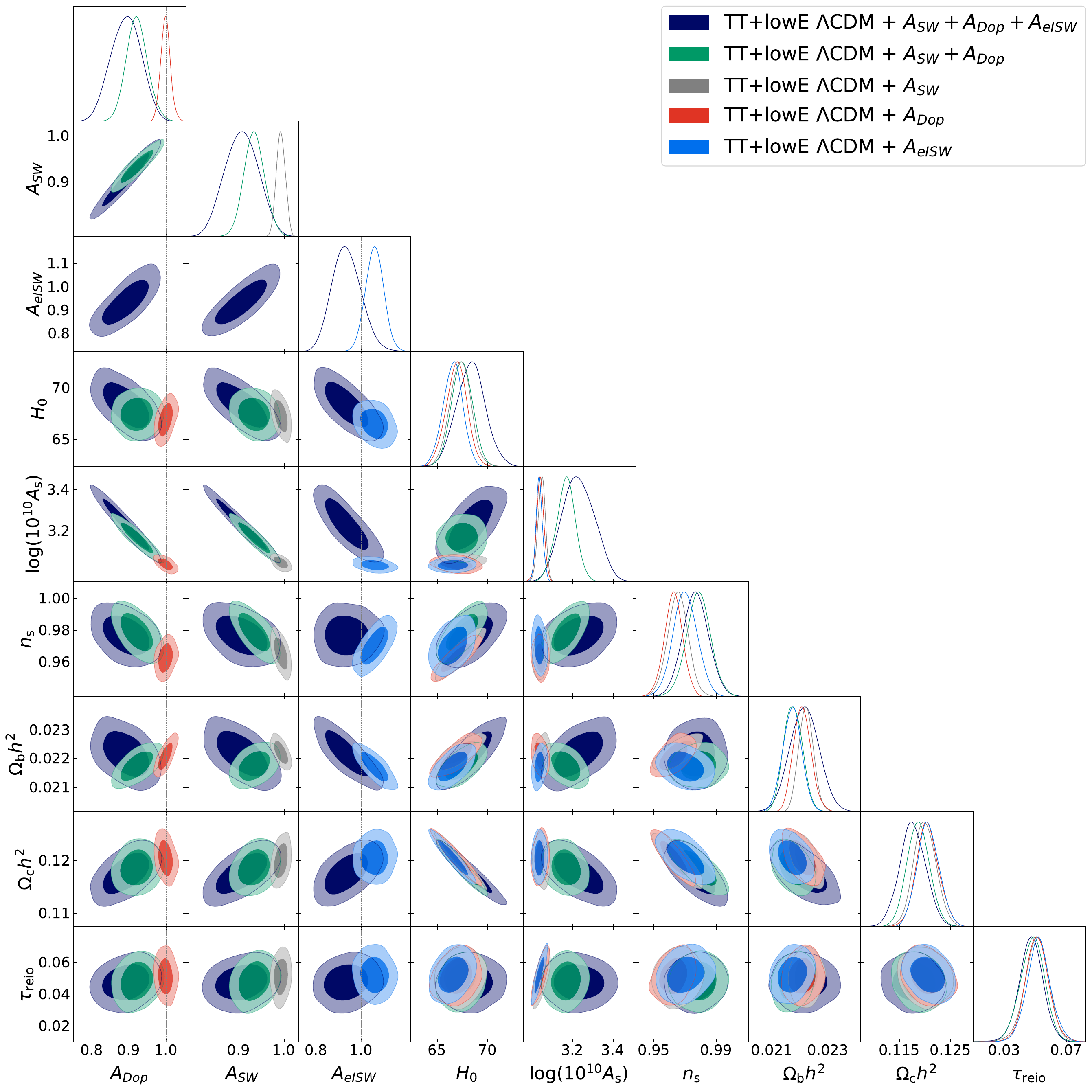}
\caption{\textit{\footnotesize{Plots of the marginalized posterior comparison of $\Lambda$CDM + $A_{SW}+A_{Dop}+A_{eISW}$, $\Lambda$CDM + $A_{SW}+A_{Dop}$, $\Lambda$CDM + $A_{SW}$, $\Lambda$CDM + $A_{Dop}$ and $\Lambda$CDM + $A_{eISW}$ fits to Planck 2018 TT+lowE data. The contours display the 68\% and 95\% limits and the black dotted lines represent the values for each phenomenological amplitude, $A_{SW}=A_{Dop}=A_{eISW}=1$, predicted by $\Lambda$CDM model.}}}\label{TTearlyISWDopSWCombination}
\end{center}
\end{figure}

\begin{figure}[p]
\centering
\includegraphics[width=1.0\textwidth]{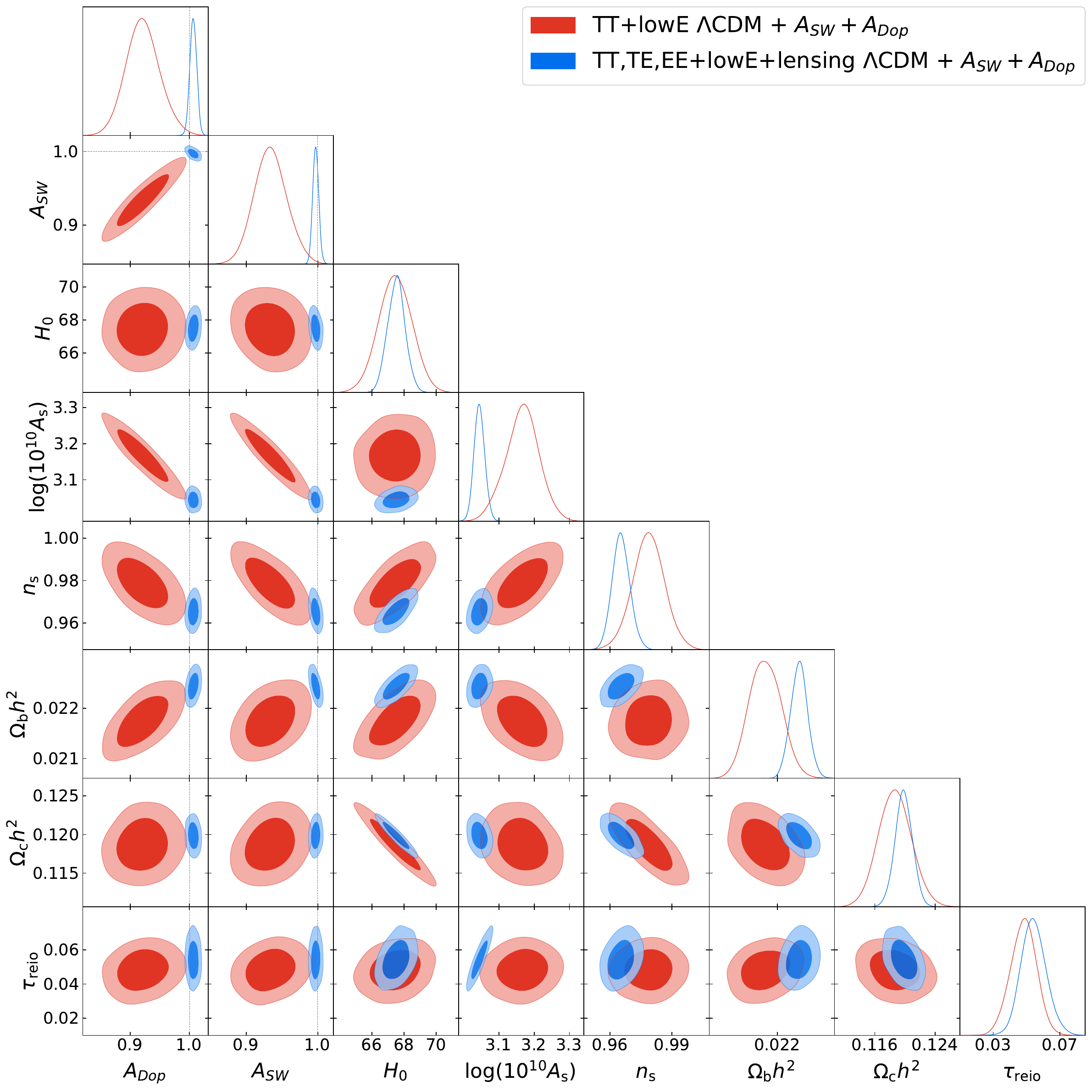}
\caption{\textit{\footnotesize{Plots of the marginalized posterior comparison of $\Lambda$CDM + $A_{SW}+A_{Dop}$ fits to Planck 2018 TT+lowE data and Planck 2018 TT,TE,EE+lowE+lensing data. The contours display the 68\% and 95\% limits and the black dotted lines represent the value for each phenomenological amplitude, $A_{SW}=A_{Dop}=1$, predicted by $\Lambda$CDM model.}}}\label{TTandPolarizationComparisonDopSW}
\end{figure}

\begin{figure}[ht]
\begin{center}
\includegraphics[width=1.0\textwidth]{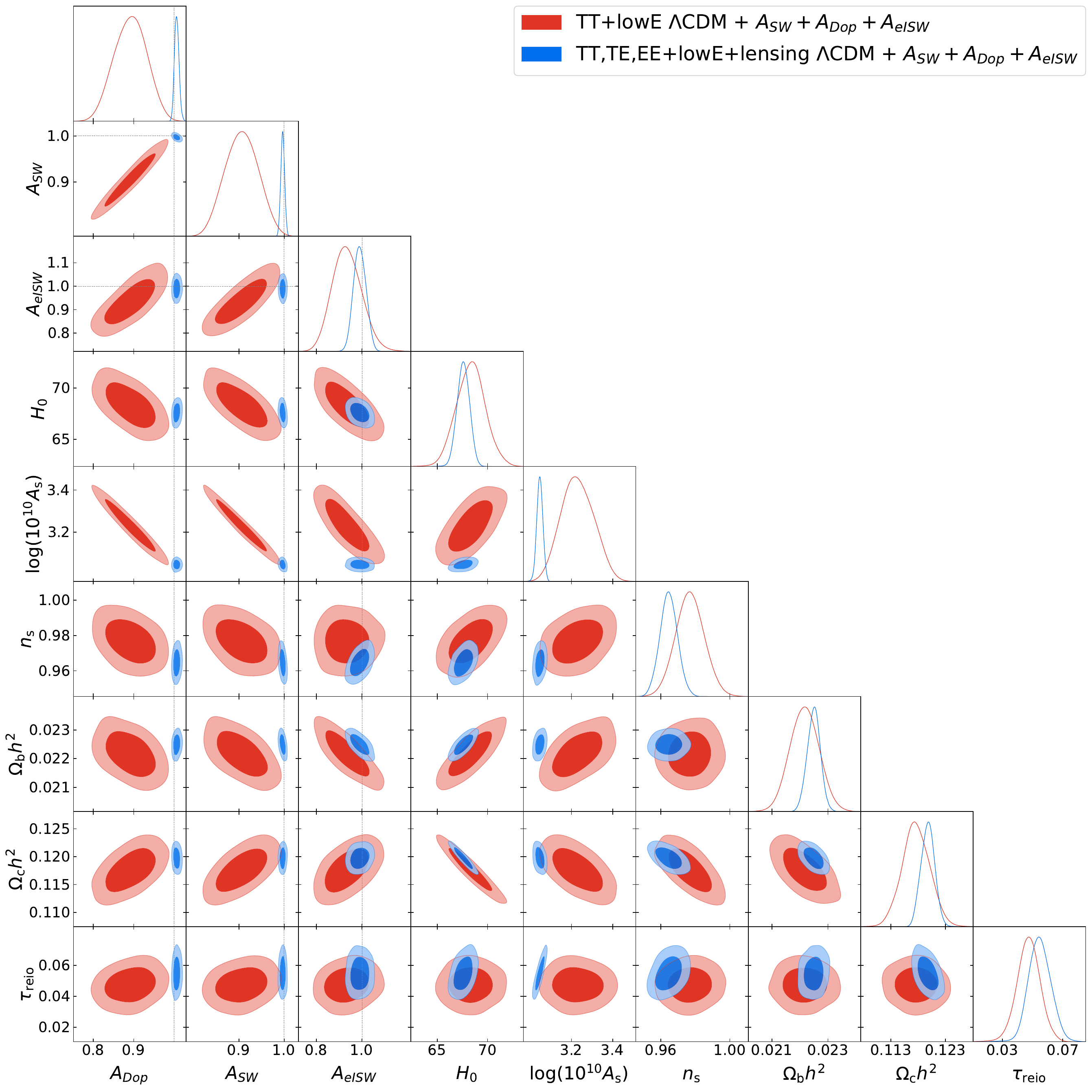}
\caption{\textit{\footnotesize{Plots of the marginalized posterior comparison of $\Lambda$CDM + $A_{SW}+A_{Dop}+A_{eISW}$ fits to Planck 2018 TT+lowE data and Planck 2018 TT,TE,EE+lowE+lensing data. The contours display the 68\% and 95\% limits and the black dotted lines represent the value for each phenomenological amplitude, $A_{SW}=A_{Dop}=A_{eISW}=1$, predicted by $\Lambda$CDM model.}}}\label{TTandPolarizationComparisonearlyISWDopSW}
\end{center}
\end{figure}

In this subsection, we analyse combinations of two and three phenomenological amplitudes in addition to the cosmological parameters of $\Lambda$CDM. First, we test all combinations of the cosmological parameters plus two phenomenological amplitudes, without considering the late ISW effect which, as explained in the previous subsection, is not constrained. In addition, we have explored a combination of three phenomenological amplitudes, $A_{SW}+A_{Dop}+A_{eISW}$, to probe the power of using polarization and lensing data to break degeneracies between parameters. The results can be seen in Tables \ref{tab:TTtwoParams} and \ref{tab:TTPoltwoParams}, and in Figures \ref{TTearlyISWDopSWCombination}, \ref{TTandPolarizationComparisonDopSW} and \ref{TTandPolarizationComparisonearlyISWDopSW}.

In Table \ref{tab:TTtwoParams} eleven different models are fitted to the Planck 2018 TT+lowE data. We will focus on the models that yield the largest improvements compared to $\Lambda$CDM. First, all models, including $A_L$, show an improvement over the $\Lambda$CDM model. Other interesting models are $A_{SW}+A_{Dop}$ and $A_{SW}+A_{Dop}+A_{eISW}$, which also improves the fit of $\Lambda$CDM. In particular, when considering the individual phenomenological amplitudes, no such good fits were obtained. It is precisely the combination of these amplitudes that allows for a better fit. In addition, the measured mean values for the phenomenological amplitudes are shifted from the expected values, e.g., for $A_{SW}$ in the $\Lambda$CDM+$A_{SW}+A_{Dop}$ model a deviation of $2.9 \sigma$ from 1 was observed. 

A question arises: is this an indication of new physics or the presence of some unknown systematics? The answer is neither, and the reasons are explained below. When using only the TT+lowE data, from Figure \ref{TTearlyISWDopSWCombination} it can be seen that $\ln(10^{10}A_s)$ is highly anti-correlated with the three phenomenological amplitudes, $A_{SW}$, $A_{eISW}$ and $A_{Dop}$. This anti-correlation is more important when the three phenomenological amplitudes are left as free parameters simultaneously.  As explained in \cite{Kable_2020}, the increase of these three phenomenological amplitudes ($A_{SW}$, $A_{eISW}$ and $A_{Dop}$) has a similar effect to the increase of $A_s$ in the TT angular power spectrum. This explains why $\ln(10^{10}A_s)$ is slightly above the value obtained in the $\Lambda$CDM fit and the phenomenological amplitudes $A_{SW}$, $A_{eISW}$ and $A_{Dop}$ are below their expected value, similarly all three. To understand this result, it must be taken into account that the late ISW effect is only of importance at low $\ell$  and the effect of the polarization contribution has very little impact on the final TT angular power spectrum, so both effects can be neglected. For this reason, the temperature angular power spectrum is roughly modulated by a product of the scalar amplitude, $A_s$, and a single phenomenological amplitude that rescales all physical contributions of the temperature transfer function \eqref{transferFunction}. Consequently, the combination of $A_{SW}$, $A_{eISW}$ and $A_{Dop}$ mimics the effect of $A_s$ in the TT angular power spectrum.

Therefore, we face a theoretical problem that even the best temperature CMB experiment would face. The solution is to make use of the cross-correlation with other data sets. When the polarization and lensing CMB data are included, the degeneracy breaks down completely. Then, the expected value of one is recovered, as can be seen in Figures \ref{TTandPolarizationComparisonDopSW} and \ref{TTandPolarizationComparisonearlyISWDopSW}, which is one of the most interesting results of this work. The reason is that the angular power spectrum of EE does not depend on the Sachs-Wolfe, the early ISW and the Doppler effects, as deduced from the equations \eqref{clTheory} and \eqref{transferFunction}, so that $A_s$ can be fully determined. Note that although the Table \ref{tab:TTtwoParams} contains polarization information (lowE), there are not enough data points to break the degeneracies between the parameters.

Returning to Table \ref{tab:TTPoltwoParams}, the combinations of $A_{SW}+A_{Dop}$ and $A_{SW}+A_{Dop}+A_{eISW}$ are not favoured models. Instead, the most favoured combination of parameters is $A_L+A_{eISW}$. However, the improvement observed in comparison to the model with only $A_L$ in Table \ref{tab:TTPolOnePhenom} is not significant.

For the rest of the models analysed in Tables \ref{tab:TTtwoParams} and \ref{tab:TTPoltwoParams}, nothing unusual was found. No important deviations of the mean values nor significant improvements to the $\Lambda$CDM model were observed. For that reason, no other combinations of three phenomenological amplitudes were studied.

\clearpage

 \begin{sidewaystable}
  \centering
   \resizebox{\columnwidth}{!}{
    \begin{tabular}{|c|c|c|c|c|c|c|c|c|c|c|c|}
    \hline
    Param. & $A_{Dop}+A_{Pol}$ & $A_{SW}+A_{eISW}$ & $A_{L}+A_{Dop}$ & $A_{L}+A_{SW}$ & $A_{L}+A_{eISW}$ & $A_{L}+A_{Pol}$ & $A_{eISW}+A_{Pol}$ & $A_{Dop}+A_{eISW}$ & $A_{SW}+A_{Pol}$ & $+A_{SW}  + A_{Dop}$ & $+A_{SW}  + A_{Dop} + A_{eISW}$ \\
    \hline\hline
    $n_s$ & $0.9625\pm 0.0065$ & $0.9698\pm 0.0075$ & $0.9734\pm 0.0072$ & $0.9726\pm 0.0071$ & $0.9743\pm 0.0075$ & $0.9743\pm 0.0079$ & $0.9705\pm 0.0086$ & $0.9698\pm 0.0083$ & $0.9647\pm 0.0072$ & $0.9787\pm 0.0077$ & $0.9769\pm 0.0083$ \\
    \hline
    $H_0$ & $66.8^{+1.2}_{-1.0}$ & $66.4\pm 1.2$ & $68.7\pm 1.2$ & $69.0\pm 1.2$ & $68.8^{+1.3}_{-1.4}$ & $69.1\pm 1.3$ & $66.6\pm 1.1$ & $66.2\pm 1.1$ & $67.2\pm 1.1$ & $67.5\pm 1.0$ & $68.3\pm 1.4$ \\
    \hline
    $100\Omega_bh^2$ & $2.203\pm 0.032$ & $2.167\pm 0.047$ & $2.240\pm 0.033$ & $2.256\pm 0.028$ & $2.252^{+0.046}_{-0.053}$ & $2.261^{+0.031}_{-0.035}$ & $2.172\pm 0.037$ & $2.154\pm 0.045$ & $2.219\pm 0.029$ & $2.175\pm 0.033$ & $2.216\pm 0.051$ \\
    \hline
    $\Omega_ch^2$ & $0.1204\pm 0.0024$ & $0.1209\pm 0.0022$ & $0.1167\pm 0.0025$ & $0.1163\pm 0.0025$ & $0.1168\pm 0.0026$ & $0.1166\pm 0.0025$ & $0.1206\pm 0.0021$ & $0.1210\pm 0.0023$ & $0.1199\pm 0.0022$ & $0.1187\pm 0.0022$ & $0.1176\pm 0.0025$ \\
    \hline
    $\tau_{reio}$ & $0.0563^{+0.0063}_{-0.015}$ & $0.0515\pm 0.0080$ & $0.0495^{+0.0094}_{-0.0073}$ & $0.0504^{+0.0087}_{-0.0077}$ & $0.0503^{+0.0094}_{-0.0078}$ & $0.0497^{+0.0091}_{-0.011}$ & $0.0521^{+0.0086}_{-0.012}$ & $0.0516\pm 0.0078$ & $0.0539^{+0.0086}_{-0.012}$ & $0.0482\pm 0.0078$ & $0.0473\pm 0.0077$ \\
    \hline
    $\ln (10^{10}A_s)$ & $3.050^{+0.016}_{-0.030}$ & $3.033\pm 0.022$ & $3.034^{+0.021}_{-0.018}$ & $3.010\pm 0.025$ & $3.027^{+0.019}_{-0.016}$ & $3.026^{+0.019}_{-0.023}$ & $3.034^{+0.020}_{-0.024}$ & $3.039\pm 0.019$ & $3.052^{+0.021}_{-0.026}$ & $3.166^{+0.049}_{-0.044}$ & $3.229\pm 0.078$ \\
    \hline
    $A_{SW}$ & $-$   & $1.000\pm 0.011$ & $-$   & $1.013\pm 0.013$ & $-$   & $-$   & $-$   & $-$   & $0.9932\pm 0.0096$ & $0.934\pm 0.023$ & $0.906\pm 0.036$ \\
    \hline
    $A_{Dop}$ & $0.997\pm 0.013$ & $-$   & $0.987\pm 0.013$ & $-$   & $-$   & $-$   & $-$   & $0.992\pm 0.013$ & $-$   & $0.921^{+0.027}_{-0.030}$ & $0.892\pm 0.040$ \\
    \hline
    $A_{eISW}$ & $-$   & $1.062\pm 0.046$ & $-$   & $-$   & $1.010\pm 0.045$ & $-$   & $1.059\pm 0.043$ & $1.065\pm 0.044$ & $-$   & $-$   & $0.935^{+0.059}_{-0.068}$ \\
    \hline
    $A_{Pol}$ & $0.96^{+0.23}_{-0.19}$ & $-$   & $-$   & $-$   & $-$   & $1.04\pm 0.22$ & $1.02\pm 0.21$ & $-$   & $0.96\pm 0.20$ & $-$   & $-$ \\
    \hline
    $A_{L}$ & $-$   & $-$   & $1.262^{+0.095}_{-0.11}$ & $1.32\pm 0.12$ & $1.227^{+0.094}_{-0.11}$ & $1.238\pm 0.097$ & $-$   & $-$   & $-$   & $-$   & $-$ \\
    \hline\hline
    $\chi^2$ & 1172.88 & 1171.02 & 1166.29 & 1166.53 & 1167.50 & 1167.27 & 1171.40 & 1168.41 & 1173.54 & 1167.35 & 1166.06 \\
    \hline
    $\chi^2_{\Lambda CDM} - \chi^2$ & -0.70  & 1.16  & 5.89  & 5.65  & 4.68  & 4.91  & 0.78  & 3.77  & -1.36 & 4.83  & 6.12 \\
    \hline
    \end{tabular}%
    }
    \caption{Mean values and 68\% confidence intervals for the minimal $\Lambda$CDM plus two phenomenological parameters (not including $A_{lISW}$) and for $\Lambda$CDM plus $A_{SW}$, $A_{Dop}$ and $A_{eISW}$, for the MCMC chains fit to Planck 2018 TT+lowE data.}\label{tab:TTtwoParams}
\vspace{1.0cm}
   \resizebox{\columnwidth}{!}{
    \begin{tabular}{|c|c|c|c|c|c|c|c|c|c|c|c|}
    \hline
    Param. & $A_{Dop}+A_{Pol}$ & $A_{SW}+A_{eISW}$ & $A_{L}+A_{Dop}$ & $A_{L}+A_{SW}$ & $A_{L}+A_{eISW}$ & $A_{L}+A_{Pol}$ & $A_{eISW}+A_{Pol}$ & $A_{Dop}+A_{eISW}$ & $A_{SW}+A_{Pol}$ & $+A_{SW}  + A_{Dop}$ & $+A_{SW}  + A_{Dop} + A_{eISW}$ \\
    \hline\hline
    $n_s$ & $0.9647\pm 0.0041$ & $0.9639\pm 0.0053$ & $0.9691\pm 0.0046$ & $0.9695\pm 0.0050$ & $0.9684\pm 0.0054$ & $0.9691\pm 0.0048$ & $0.9627\pm 0.0051$ & $0.9631\pm 0.0048$ & $0.9674\pm 0.0046$ & $0.9654\pm 0.0043$ & $0.9646\pm 0.0051$ \\
    \hline
    $H_0$ & $67.42\pm 0.55$ & $67.42\pm 0.57$ & $68.21^{+0.71}_{-0.63}$ & $68.16\pm 0.70$ & $68.25\pm 0.72  $ & $68.13\pm 0.69$ & $67.38^{+0.52}_{-0.58}$ & $67.46\pm 0.56$ & $67.67\pm 0.58$ & $67.52\pm 0.55$ & $67.59\pm 0.61$ \\
    \hline
    $100\Omega_bh^2$ & $2.241\pm 0.017$ & $2.243\pm 0.022$ & $2.252\pm 0.016$ & $2.250\pm 0.018$ & $2.257\pm 0.023$ & $2.248\pm 0.017$ & $2.239\pm 0.020$ & $2.243\pm 0.020$ & $2.246\pm 0.018$ & $2.244\pm 0.017$ & $2.249\pm 0.023$ \\
    \hline
    $\Omega_ch^2$ & $0.1200\pm 0.0012$ & $0.1200\pm 0.0012$ & $0.1182^{+0.0014}_{-0.0015}$ & $0.1182\pm 0.0015$ &  $0.1182\pm 0.0015$ & $0.1183^{+0.0014}_{-0.0016}$ & $0.1200\pm 0.0012$ & $0.1199\pm 0.0012$ & $0.1195\pm 0.0012$ & $0.1198\pm 0.0012$ & $0.1198\pm 0.0013$ \\
    \hline
    $\tau_{reio}$ & $0.0541\pm 0.0074$ & $0.0538\pm 0.0073$ & $0.0490^{+0.0082}_{-0.0066}$ & $0.0494^{+0.0087}_{-0.0076}$ & $0.0491^{+0.0078}_{-0.0071}$ & $0.0487^{+0.0084}_{-0.0075}$ & $0.0539^{+0.0065}_{-0.0076}$ & $0.0537\pm 0.0071$ & $0.0532\pm 0.0077$ & $0.0543^{+0.0069}_{-0.0079}$ & $0.0548^{+0.0068}_{-0.0076}$ \\
    \hline
    $\ln (10^{10}A_s)$ & $3.040\pm 0.015$ & $3.045\pm 0.015$ & $3.026^{+0.017}_{-0.015}$ & $3.029\pm 0.018$ & $3.030\pm 0.016$ & $3.028^{+0.018}_{-0.016}$ & $3.045^{+0.013}_{-0.015}$ & $3.041\pm 0.014$ & $3.056\pm 0.017$ & $3.044\pm 0.015$ & $3.046\pm 0.015$ \\
    \hline
    $A_{SW}$ & $-$   & $0.9992\pm 0.0038$ & $-$   & $0.9999\pm 0.0037$ & $-$   & $-$   & $-$   & $-$   & $0.9887\pm 0.0079$ & $0.9971\pm 0.0042$ & $0.9969\pm 0.0042$ \\
    \hline
    $A_{Dop}$ & $1.0068\pm 0.0083$ & $-$   & $1.0053\pm 0.0049$ & $-$   & $-$   & $-$   & $-$   & $1.0047\pm 0.0051$ & $-$   & $1.0065\pm 0.0058$ & $1.0065^{+0.0052}_{-0.0059}$ \\
    \hline
    $A_{eISW}$ & $-$   & $0.989\pm 0.025$ & $-$   & $-$   & $0.988\pm 0.026 $ & $-$   & $0.987\pm 0.026$ & $0.988\pm 0.026$ & $-$   & $-$   & $0.990\pm 0.027$ \\
    \hline
    $A_{Pol}$ & $1.0014\pm 0.0041$ & $-$   & $-$   & $-$   & $-$   & $0.9991\pm 0.0024$ & $0.9985\pm 0.0025$ & $-$   & $0.9920\pm 0.0052$ & $-$   & $-$ \\
    \hline
    $A_{L}$ & $-$   & $-$   & $1.073\pm 0.040$ & $1.071\pm 0.041$ & $1.072\pm 0.039 $ & $1.071\pm 0.039$ & $-$   & $-$   & $-$   & $-$   & $-$ \\
    \hline \hline
    $\chi^2$ & 2764.92 & 2764.81 & 2763.75 & 2762.39 & 2761.65 & 2764.01 & 2767.39 & 2765.20 & 2764.34 & 2764.94 & 2766.17 \\
    \hline
    $\chi^2_{\Lambda CDM} - \chi^2$ & 1.89  & 2.00     & 3.06  & 4.42  & 5.16  & 2.80   & -0.58 & 1.61  & 2.47  & 1.87  & 0.64 \\
    \hline
    \end{tabular}%
    }
    \caption{Mean values and 68\% confidence intervals for the minimal $\Lambda$CDM parameters plus two phenomenological parameters (not including $A_{lISW}$) and for $\Lambda$CDM plus $A_{SW}$, $A_{Dop}$ and $A_{eISW}$, for the MCMC chains fit to Planck 2018 TT,TE,EE+lowE+lensing data.}\label{tab:TTPoltwoParams}
\end{sidewaystable}

\clearpage

\clearpage
\subsection{Four phenomenological amplitudes}

\begin{table}[htbp]
  \centering
  \resizebox{0.8\columnwidth}{!}{
    \begin{tabular}{|c|c|c|c|}
    \hline
    Param. & $\Lambda$CDM & $+A_{SW}  + A_{Dop} + A_{eISW} + A_L$ & $+A_{SW}  + A_{Dop} + A_{eISW} + A_{Pol}$  \\
    \hline \hline
    $n_s$ & $0.9645\pm 0.0041$ & $0.9682\pm 0.0055$ & $0.9681\pm 0.0056$  \\
    \hline
    $H_0$ & $67.30\pm 0.54$ & $68.34\pm 0.71$ & $68.30\pm 0.76$  \\
    \hline
    $100\Omega_bh^2$ & $2.235\pm 0.014$ & $2.263\pm 0.025$ & $2.263\pm 0.025$  \\
    \hline
    $\Omega_ch^2$ & $0.1201\pm 0.0012$ & $0.1181\pm 0.0015$ & $0.1182\pm 0.0016$  \\
    \hline
    $\tau_{reio}$ & $0.0536^{+0.0065}_{-0.0074}$ & $0.0488^{+0.0088}_{-0.0069}$ & $0.0496\pm 0.0078$  \\
    \hline
    $\ln (10^{10}A_s)$ & $3.044\pm 0.014$ & $3.028^{+0.019}_{-0.016}$ & $3.092\pm 0.034$  \\
    \hline
    $A_{SW}$ & $-$   & $0.9981\pm 0.0042$ & $0.968\pm 0.019$  \\
    \hline
    $A_{Dop}$ & $-$   & $1.0061\pm 0.0053$ & $0.976\pm 0.020$  \\
    \hline
    $A_{eISW}$ & $-$   & $0.987\pm 0.028$ & $0.958\pm 0.033$  \\
    \hline
    $A_{L}$ & $-$   & $1.073\pm 0.040$ & $-$  \\
    \hline
    $A_{Pol}$ & $-$   & $-$   & $0.970\pm 0.020$  \\
    \hline \hline
    $\chi^2$ & $2766.81$ & $2762.80$ & $2764.80$  \\
    \hline
    $\chi^2_{\Lambda CDM} - \chi^2$ & 0.00     & 4.01   & 2.01  \\
    \hline
    \end{tabular}%
    }
    \caption{Mean values and 68\% confidence intervals for the minimal $\Lambda$CDM parameters plus two combinations of four phenomenological amplitudes, for the MCMC chains fit to Planck 2018 TT,TE,EE+lowE+lensing data.}
  \label{tab:fourParams}%
\end{table}%

\begin{figure}[ht]
\begin{center}
\includegraphics[width=1.0\textwidth]{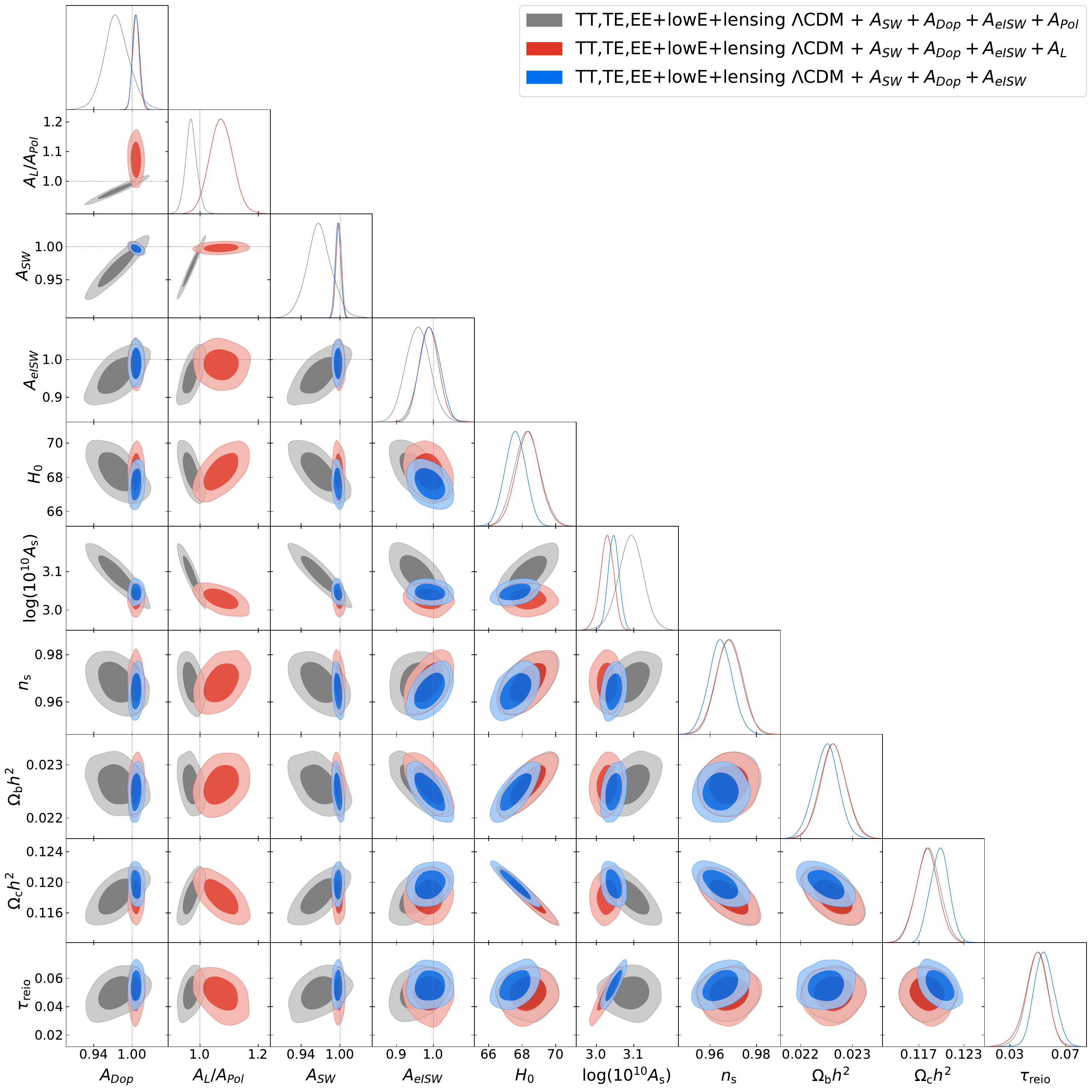}
\caption{\textit{\footnotesize{Plots of the marginalized posterior comparison of $\Lambda$CDM + $A_{SW}+A_{Dop}+A_{eISW}+A_{Pol}$ and $\Lambda$CDM + $A_{SW}+A_{Dop}+A_{eISW}+A_{L}$ fits to Planck 2018 TT,TE,EE+lowE+lensing data. The contours display the 68\% and 95\% limits and the black dotted lines represent the value for each phenomenological amplitude predicted by $\Lambda$CDM model. The parameter $A_L/A_{Pol}$ correspond to either the $A_L$ or to the $A_{Pol}$ phenomenological amplitude.}}}\label{TTTEEEDopSWeISWLensPol}
\end{center}
\end{figure}

In this subsection, two models with four phenomenological amplitudes were studied taking into account the full Planck 2018 likelihood. Planck 2018 TT+lowE data were not fitted due to the degeneracy between parameters observed in section \ref{twoThreeParamsSection}, which only could get worse when including more parameters. The objective of this section is to test the limits in which, using all Planck CMB data, the phenomenological parameters, i.e., physical processes could be well constrained. This is an additional advantage of using polarization and lensing data and not only the temperature spectrum. The results are presented in Table \ref{tab:fourParams} and Figure \ref{TTTEEEDopSWeISWLensPol}. 

The model $\Lambda\text{CDM}+A_{SW}  + A_{Dop} + A_{eISW} + A_L$ could be well constrained and no additional comments need to be made, however for the model $\Lambda\text{CDM}+A_{SW}  + A_{Dop} + A_{eISW} + A_{Pol}$ degeneracies between parameters appear involving the scalar amplitude $A_s$. The existing anti-correlation between $A_{SW} + A_{Dop} + A_{eISW}$ and $A_s$ was broken when the polarization and lensing angular power spectra are included, so how are we in the same situation again? The Figure \ref{TTTEEEDopSWeISWLensPol} and the Table \ref{tab:fourParams} have the clue to what is going on here. 

When the excluded parameter is the lensing effect, strong anticorrelations are observed between the phenomenological amplitudes $A_{SW}$, $A_{eISW}$, $A_{Dop}$ and $A_{Pol}$ and the scalar amplitude $\ln(10^{10}A_s)$. Interestingly, $A_{SW}$, $A_{eISW}$, $A_{Dop}$ were not anti-correlated with $A_s$ because the angular power spectrum of EE helps to constrain the scalar amplitude since the EE spectrum does not depend on the parameters $A_{SW}$, $A_{eISW}$ and $A_{Dop}$. The same is true for $A_s$ and $A_{Pol}$. These two parameters are extremely anti-correlated in the EE angular power spectrum, but when the temperature angular power spectrum is considered, the degeneracy disappears. The combined action of these four phenomenological amplitudes causes the existing anti-correlation with $A_s$. Note that the lensing angular power spectrum is proportional to $A_s$ and $A_L$, however, it is not sufficient to determine $A_s$ accurately and break the anti-correlations between certain parameters. It would be necessary to add other data sets to accurately constrain some combinations of four cosmological amplitudes and the combination of all phenomenological amplitudes, which is not contemplated in this work.

Finally, from Table \ref{tab:fourParams} no improvements were observed for the models studied in this section with respect to $\Lambda$CDM.

\clearpage
\section{Conclusions} \label{sec:conclusions}

In this work, we introduce phenomenological amplitudes to account for different physical contributions to the CMB angular power spectra and test how well the Planck 2018 data can constrain different combinations of cosmological parameters and phenomenological amplitudes. To do so, we adapted the \texttt{CLASS} code to include the new parameters in the calculation of the CMB angular power spectra and used the MCMC sampler of \texttt{Cobaya} to fit the Planck 2018 likelihood to different models. This work is relevant as it is the first time that a comprehensive analysis using Planck 2018 temperature, polarization and lensing data is used to study phenomenologically the Sachs-Wolfe, early and late Integrated Sachs-Wolfe, Doppler, polarization contribution and lensing effects.

The results obtained in this paper show that the constraints imposed on the phenomenological parameters are considerably tighter when using temperature, polarization and lensing data instead of using only the TT spectrum as in \cite{Kable_2020}. Furthermore, the inclusion of temperature, polarization and lensing data allows us to break the degeneracy between the three parameters $A_{SW}$, $A_{Dop}$ and $A_{eISW}$, and $A_s$, which arises when only the temperature angular power spectrum is considered. This allows us to explore combinations of four cosmological amplitudes, which is complicated with only the TT spectrum. However, the analysis of certain models with four and five phenomenological amplitudes would require the use of additional cosmological data from non-CMB experiments, which is not done in this work. 

Finally, another important feature of this article is the consistency checks of the $\Lambda$CDM model with extra-parametrisations. No inconsistencies were found and the greatest improvements were obtained for models including the lensing parameter, $A_L$, which is well-known from the Planck Collaboration analyses.


\acknowledgments
The authors would like to thank Julien Lesgourges and Jesús Torrado for helpful comments on the modifications introduced in \texttt{CLASS} and the usage of \texttt{Cobaya}. MRG acknowledges financial support from a Collaboration Grant with the Modern Physics Department at the Universidad de Cantabria from the Spanish Ministry of Education, Culture and Sports and from a JAE Intro 2021 grant from the Spanish National Research Council (CSIC).
PV thanks Spanish Agencia Estatal de Investigaci\'on (AEI, MICIU) for the financial support provided under the projects with references  PID2019-110610RB-C21 and ESP2017-83921-C2-1-R, and also acknowledge the funding from Unidad de Excelencia Mar{\'\i}a de Maeztu (MDM-2017-0765). The authors acknowledge the computer resources, technical expertise and assistance provided by the Spanish Supercomputing Network (RES) node at Universidad de Cantabria. 
We make use of the \texttt{numpy} \cite{numpy} and \texttt{matplotlib} \cite{matplotlib} Python packages in some calculations and figures. For all the computations using \texttt{CLASS}, the non-linear corrections were included using the software \texttt{HMCode} \cite{HMCode}.

\bibliographystyle{JHEP}
\bibliography{Bibliography}
\end{document}